\documentclass[conference]{IEEEtran}
\IEEEoverridecommandlockouts
\usepackage{cite}
\usepackage{amsmath,amssymb,amsfonts}
\usepackage{algorithmic}
\usepackage{graphicx}
\usepackage{textcomp}
\usepackage{xcolor}
\usepackage[utf8]{inputenc}
\usepackage{textgreek}
\usepackage{subcaption}
\def\BibTeX{{\rm B\kern-.05em{\sc i\kern-.025em b}\kern-.08em
    T\kern-.1667em\lower.7ex\hbox{E}\kern-.125emX}}

\makeatletter
\newcommand{\linebreakand}{%
    \end{@IEEEauthorhalign}
    \hfill\mbox{}\par
    \mbox{}\hfill\begin{@IEEEauthorhalign}
}

\begin{document}
\title{Implementation of the Spherical Trigonometry Algorithm in a Qibla Direction Application\\}

\author{\IEEEauthorblockN{1\textsuperscript{st} Wisnu Uriawan}
\IEEEauthorblockA{\textit{Informatics Department}\\
\textit{UIN Sunan Gunung Djati Bandung}\\
Jawa Barat, Indonesia\\
wisnu\_u@uinsgd.ac.id}
\and
\IEEEauthorblockN{2\textsuperscript{nd} Andika Nuralamsyah}
\IEEEauthorblockA{\textit{Informatics Department}\\
\textit{UIN Sunan Gunung Djati Bandung}\\
Jawa Barat, Indonesia\\
syahandikanuralam@gmail.com}
\and
\IEEEauthorblockN{3\textsuperscript{rd} Abidzar Giffari}
\IEEEauthorblockA{\textit{Informatics Department}\\
\textit{UIN Sunan Gunung Djati Bandung}\\
Jawa Barat, Indonesia\\
abidzar.fathoni321@gmail.com}
\linebreakand
\IEEEauthorblockN{4\textsuperscript{th} Arief Rahman Mubarok}
\IEEEauthorblockA{\textit{Informatics Department}\\
\textit{UIN Sunan Gunung Djati Bandung}\\
Jawa Barat, Indonesia\\
ariefmubarok25@gmail.com}
\and
\IEEEauthorblockN{5\textsuperscript{th} Aisyah Muthmainnah}
\IEEEauthorblockA{\textit{Informatics Department}\\
\textit{UIN Sunan Gunung Djati Bandung}\\
Jawa Barat, Indonesia\\
aisyaaahm2506@gmail.com}
\and
\IEEEauthorblockN{6\textsuperscript{th} Najwa Naura Salsabilla Herdiyana}
\IEEEauthorblockA{\textit{Informatics Department}\\
\textit{UIN Sunan Gunung Djati Bandung}\\
Jawa Barat, Indonesia\\
salsabillaherdiyana@gmail.com}
}

\maketitle

\begin{abstract}
This research explores the implementation of a spherical trigonometry algorithm in the development of QiblatKita, a mobile-based Qibla direction application designed to address the need for accurate and accessible prayer direction tools for Muslims, especially in the context of global mobility and diverse geographical settings. Built using the Flutter framework to ensure cross-platform compatibility and near-native performance, the application integrates critical device sensors—GPS, magnetometer, and accelerometer—to collect real-time positional and orientation data. The Agile Software Development Lifecycle (SDLC) was adopted to facilitate iterative refinement and responsiveness to user feedback. By applying the Haversine formula and Spherical Law of Cosines, the system calculates the azimuth angle between the user’s location and the Ka’aba with high precision, achieving an angular deviation of less than one degree in testing. The results confirm that QiblatKita offers a stable, user-friendly, and scientifically-grounded solution, effectively merging spherical trigonometry with modern mobile technology to support religious practice in the digital era.
\end{abstract}


\section{Introduction} \label{sec:introduction}

In the digital era characterized by rapid technological progress, smartphones have evolved into multifunctional devices that support various aspects of daily life—including religious practices. For Muslims around the world, one of the fundamental daily obligations is to perform the five obligatory prayers, which require facing the Qibla, the direction of the Kaaba located in Masjid al-Haram, Mecca. Given the global distribution of Muslim populations and the increasingly mobile nature of modern life, determining the precise direction of Qibla in unfamiliar locations has become a practical challenge  \cite{mahmud2022penentuan}.This classical problem has found its modern solution through the integration of Islamic scholarship and computer science.

This solution is manifested in the development of digital Qibla compass applications, which can be installed on mobile devices and provide convenience through embedded smartphone sensors. Behind their simple and user-friendly interfaces lies a sophisticated computational process. The intelligence of these applications lies in their ability to solve geometric problems on the Earth’s surface—naturally a spherical body. At this point, the spherical trigonometric algorithms play a crucial role as the mathematical core that enables accurate Qibla direction determination.

In today’s digital landscape, mobile application marketplaces such as the Play Store are saturated with software based on modern Islamic astronomy (falak). These applications are designed with intuitive interfaces and highly user-oriented operational processes, making them accessible to people from diverse backgrounds, even those without a deep understanding of Islamic astronomy. Such accessibility and practicality have driven significant user interest among Muslims to download and adopt these applications on their mobile devices \cite{sriani2022uji}\cite{aurelia2022akurasi}. This phenomenon represents a paradigm shift in how religious obligations are fulfilled—technology now serves as an efficient and instant facilitator. Rather than relying on complex manual calculations or consulting experts, modern Muslims increasingly depend on digital applications to support their religious routines, such as determining prayer times, Qibla direction, and imsakiyah schedules during Ramadan, thereby simplifying and enhancing their daily worship practices.

The Kaaba, located in Masjid al-Haram in Mecca, serves as the sacred focal point for Muslims worldwide. Its role as a center of worship dates back to the Prophet Ibrahim (Abraham) and his son Ismail, who built it under the command of Allah Subhanahu wa Ta’ala \cite{al2015sejarah}. The Qur’an explicitly mentions this in Surah Al-Baqarah (2:144): “We have certainly seen the turning of your face toward the heaven, and We will surely turn you to a Qibla with which you will be pleased. So turn your face toward al-Masjid al-Haram.” This verse marks not only the geographic redirection of prayer from Bayt al-Maqdis to the Kaaba but also a profound reaffirmation of monotheistic faith and spiritual identity.

Every Muslim performing prayer, regardless of location, is obligated to face the Qibla. This obligation constitutes one of the essential conditions for the validity of prayer, uniting Muslims from Indonesia to Morocco in a single direction of worship. The requirement is based on clear legal evidence from the Qur’an and Sunnah, emphasizing the necessity of facing the Kaaba (Baitullah) during prayer as a form of absolute devotion to the Creator \cite{iman2017peranan}.

The accuracy of the Qibla direction is therefore inseparable from the validity of worship itself. A Muslim is required to make a sincere effort to determine the correct direction—whether through natural indicators, magnetic compasses, or modern technologies such as GPS-based mobile applications \cite{amiral2010aplikasi}. Such precision reflects both diligence in fulfilling divine commandments and awareness that prayer represents intimate spiritual communication with Allah. Consequently, striving to face the Kaaba precisely is an expression of piety and submission that enhances the concentration and perfection of the prayer ritual.

Spherical trigonometry is a branch of mathematics that studies the relationships between angles and sides on triangles situated on the surface of a sphere. Unlike planar trigonometry, which applies to flat surfaces, spherical trigonometry is specifically designed to address curvature, making it a highly relevant mathematical framework for modeling the Earth. In this context, the geoid shape of the planet is approximated as a perfect sphere to simplify global navigation calculations, enabling accurate results for various applications—from aviation and geolocation to determining the Qibla direction—while accounting for Earth’s spherical nature.

As a mathematical discipline, spherical trigonometry provides an ideal framework for Earth modeling \cite{maftukhah2018analisis}. Its key strength lies in defining what is known as a spherical triangle formed by three points connected through great-circle arcs—the shortest paths between any two points on a sphere. In the context of spiritual navigation, these three critical points are the user’s location, the North Pole, and the Kaaba’s coordinates in Mecca, forming an imaginary spherical triangle used for Qibla calculations.

The spherical trigonometry algorithm enables accurate computation of the bearing angle from any point on Earth toward the Kaaba by utilizing latitude and longitude as primary inputs along with trigonometric functions such as sine, cosine, and tangent to determine the direction relative to true north. In its implementation, a Qibla compass application integrates GPS data to acquire user coordinates, a magnetometer to detect the Earth’s magnetic field, and an accelerometer to adjust device orientation, ensuring precision even when the device is tilted. Several computational approaches—including the Haversine formula, Spherical Law of Cosines, and vector mathematics—are applied to improve both accuracy and stability. The use of such algorithms demonstrates a synthesis between Islamic jurisprudential precision and scientific accuracy, producing digital tools that allow Muslims worldwide to identify the Qibla direction with confidence and convenience.

An innovative solution for determining the Qibla in the modern era is therefore the development of a Qibla compass application based on spherical trigonometric algorithms. This application utilizes GPS to obtain user coordinates and employs magnetometer and accelerometer sensors to determine device orientation. It then computes the Qibla direction accurately using formulas such as Haversine and Spherical Law of Cosines, which consider Earth’s curvature. The results are visualized through a digital compass interface or an augmented-reality feature that displays the Qibla direction in real time. With an intuitive interface and additional functionalities such as magnetic declination correction and favorite-location storage, the application offers precision, ease of use, and a personalized experience for Muslims seeking to determine the Qibla direction wherever they may be.

\section{Related Work} \label{sec:related-work}
In previous studies, there have been many researches discussing this topic, but there are still many shortcomings and differences. One of them is a study entitled "Trigonometry Formulation in Developing Qibla Direction Determination Software Based on Visual Basic" by Mustafa Syukur and Muhammad Hasan Basri. This research discusses the development of a Qibla direction determination application using the approach of Islamic astronomy, which is divided into ilmiy falak (theoretical) and amaliy falak (practical).This research focuses on calculating the Qibla direction based on the angle between the meridian line of a location and the great circle connecting that location to the Ka'bah, as well as determining the time when the sun is exactly on the path towards the Ka'bah. The researcher developed a solution in the form of an application program built using Visual Basic to make it easier for the general public to determine the Qibla direction without having to understand the complex calculations of astronomy. This study aims to address the public's limited knowledge in accurately calculating the Qibla direction, which until now has relied solely on estimation \cite{syukur2019formulasi}.

The main advantage of this research lies in its comprehensive approach that combines both theoretical and practical aspects of astronomy. By considering two methods simultaneously—angle calculations based on spherical trigonometry and tracking the sun's position—this application offers a more diverse range of methods for determining the qibla direction. The use of Visual Basic as a development platform was also a suitable choice at the time, given its ease in creating a user-friendly graphical interface. Furthermore, this research has high practical value because it is specifically aimed at addressing a real problem in society, namely the reliance on qibla directions that are still approximate.

However, the study has several limitations when viewed in the context of modern technology. The use of Visual Basic as a development platform restricts the application's accessibility to desktop users only, while computing trends have shifted to mobile devices. In addition, the study does not explain in detail the spherical trigonometry algorithms used in the calculation of the qibla direction. This is the main distinction from our research titled "Implementation of Spherical Trigonometry Algorithms in a Qibla Direction Guidance Application" using Flutter. Our study specifically implements spherical trigonometry algorithms such as the Haversine formula and optimizes them for mobile devices by integrating GPS, magnetometer, and accelerometer sensors.

Meanwhile, our research offers several different innovations that align with the latest technological developments. By using Flutter as the development framework, the resulting application can run natively on both iOS and Android platforms, reaching a wider range of users. The implementation of spherical trigonometry algorithms is carried out with a more modern and accurate approach, supplemented with magnetic declination correction to improve precision.

The article entitled ``The Application of Trigonometry to Determine Qibla Direction Corrections Using the Solar Azimuth Measurement Method'' by Sunanah Lailatus Sadiyah addresses a real problem faced by the students of the Al-Faraby Unit at UIN Malang, where there is a misconception that the qibla direction is identical to the west. This study uses a spherical trigonometry approach with direct solar azimuth measurement methods using a theodolite and a Rulof prism. The results indicate a correction of 2°43'33.13" is needed to adjust the qibla direction at this unit. This case study confirms that the qibla direction is specific to the geographical coordinates of each location and cannot be generalized as simply facing west \cite{sunanah2006aplikasi}.

The main advantage of this research lies in its empirical approach and high technical precision. By using a theodolite and ruling prism, the research was able to produce accurate solar azimuth data to determine the correction of the qibla direction. This direct measurement method provides strong field validation for spherical trigonometric theoretical calculations. In addition, this study successfully identified and quantified the specific qibla orientation errors in a real building, providing practical solutions that can be directly applied by the residents of the Al-Faraby Unit. This detailed case study approach provides valuable contributions to understanding the complexities of determining the direction of the Qibla at the architectural level.

Although accurate, the study has limitations in terms of accessibility and replicability. The use of a theodolite and Rulof prism requires specialized technical expertise and can only be carried out by professionals, making it inaccessible to the general public. Additionally, the method of measuring the sun's azimuth can only be performed under clear weather conditions, limiting the timing of measurements. This is a fundamental difference from our research on the "Implementation of Spherical Trigonometry Algorithms in a Qibla Direction Guidance Application" using Flutter. Our study develops a digital solution that can be accessed anytime and by anyone via a smartphone, without requiring special technical skills or specific weather conditions.

Our research offers a more practical and decentralized approach compared to the reference article. By implementing spherical trigonometry algorithms in a Flutter-based mobile application, the proposed solution can reach a wider range of users without relying on specialized tools or experts. While Sunanah et al.'s research focuses on correcting the qibla direction for a specific building, my research develops an assistive tool that can be used in various locations in real-time. Another advantage lies in my application's ability to provide dynamic qibla direction guidance, adjusting to the user's movements.

The article entitled "Questioning the Accuracy of the Qibla Direction of the Grand Mosque: A Case Study of the Grand Mosque of Parepare in the Review of Qibla Direction Trigonometry" by ABD Karim Faiz, Budiman Budiman Budiman, and Muh Rasywan Syarif examines the accuracy of the Qibla direction of the A.G.K.H. Abdul Rahman Ambo Dalle Grand Mosque in Parepare City through a multidisciplinary approach. This study is motivated by the importance of calibrating the mosque's Qibla direction in line with the development of astronomical science. Based on initial measurements using Google Earth, the research team found a deviation of the mosque's Qibla direction toward the south. This study aims to describe the methods previously used to determine the Qibla direction and to analyze and recalibrate the Qibla direction using contemporary astronomical calculations through various methods \cite{faiz2022menyoal}.

The main advantage of this research lies in its comparative approach, which utilizes three different calibration methods – Google Earth, the gnomon, and the theodolite. Cross-validation through multiple methods produces more reliable and comprehensive findings. The consistency of results from the three methods, which indicate a significant deviation (5°30' to 6°), strengthens the validity of the research findings. Furthermore, this study not only focuses on technical aspects but also includes historical analysis, thereby providing a complete understanding of the evolution of qibla determination in the mosque. This multidisciplinary approach offers a significant added value in the study of contemporary astronomical sciences.

Although comprehensive, the study has several limitations in terms of accessibility and practical implementation. The use of the theodolite and meridian staff requires specialized skills and can only be applied by professionals, so it cannot be easily replicated by the general public. Furthermore, the research focuses on calibrating the Qibla direction for a specific building only. This is the fundamental difference from our study on "Implementation of Spherical Trigonometry Algorithms in a Qibla Direction Guidance Application" using Flutter. Our research develops a digital solution that is universal and can be accessed by anyone, anytime, and anywhere through mobile devices.

Our research offers innovations that complement the findings of the referenced article. While the study by Abd Karim Faiz et al. focuses on correcting the Qibla direction of existing mosques, our research develops a preventive system through a mobile application that can be used from the planning stage. The implementation of spherical trigonometry algorithms on the Flutter platform enables accurate determination of the Qibla direction without relying on expensive and complex specialized tools. Another advantage lies in the scalability of our solution, which can reach all layers of society, not limited to professionals or specific institutions. Thus, our research transforms the concept of Qibla direction trigonometry from merely a calibration tool into a practical spiritual navigation system that is easily accessible for every Muslim in fulfilling the requirements of valid prayer.

Another article titled "Determination of the Qibla Direction of Campus Mosques in Malang City from the Perspective of Spherical Trigonometry and the 2010 Fatwa of the Indonesian Ulema Council No. 5" by Alya and Shahira examines an interesting phenomenon regarding variations in the Qibla direction at three campus mosques in Malang City. This study was motivated by the finding that mosques on Islamic campuses (UIN Malang and UNISMA) actually showed deviations in the Qibla direction, while mosques on public campuses (Brawijaya University) had a more precise Qibla direction. The study uses a dual approach, namely an empirical-qualitative approach through observation and interviews, as well as a falak (Islamic astronomy) scientific approach by applying spherical trigonometry and Google Earth. The aim is to analyze the consistency of the Qibla direction of these three mosques with the MUI Fatwa No. 5 of 2010 \cite{alya2025penentuan}.

The main strength of this study lies in its comprehensive integration of religious and scientific aspects. By combining spherical trigonometry analysis and the implementation of MUI fatwas, this study has successfully developed a holistic analytical framework. The comparative approach to three mosques with different backgrounds provides a rich perspective on the evolution of qibla direction determination methods over time. The findings linking the accuracy of the qibla direction to the year of construction and the methods used (from the istiwa' stick to the assistance of astronomical experts) make a valuable contribution to the understanding of the development of applied astronomy in Indonesia. In addition, the recommendation for a re-evaluation of the qibla direction of existing mosques demonstrates a concern for the precise implementation of sharia law.

Although comprehensive, the study has limitations in terms of practical solutions to the problems identified. The study focuses on analyzing existing conditions without offering technological solutions that can be implemented on a mass scale to prevent similar deviations in the future. In addition, the approach used is still conventional and requires special expertise in measurement. This is the fundamental difference between this study and our research on “Implementation of Spherical Trigonometry Algorithms in Qibla Direction Guidance Applications” using Flutter. Our research developed a mobile technology-based preventive solution that can be used from the planning stage to periodic monitoring, without relying directly on astronomical experts.

Our research offers a breakthrough in democratizing access to accurate Qibla direction determination. While Alya and Shahira's research focuses on evaluating the Qibla direction of existing mosques, our study develops a tool that can be used by anyone to verify the Qibla direction anytime and anywhere. The implementation of spherical trigonometry in a Flutter-based mobile application allows every Muslim to determine the Qibla direction with sufficient accuracy for practical needs, while also addressing the recommendations in MUI Fatwa No. 5 of 2010. Another advantage lies in the scalability of our solution, which can reach not only mosque construction but also individual needs in daily worship activities. Thus, our research complements the findings of the referenced articles by providing practical and easily accessible technological tools.
 
\section{Methodology} \label{sec:methodology}

The Agile Software Development Life Cycle (SDLC) is a software engineering methodology characterized by iterative and incremental processes. This approach has gained wide recognition for its practicality and relevance in contemporary system development \cite{larasati2021systematic}. The Agile development method was chosen for the development of a Qibla Compass application because of its advantages in flexibility, responsiveness to change, and focus on team collaboration \cite{uriawan2024implementing}.  Unlike traditional linear models such as the waterfall model, which tend to be rigid and sequential, Agile emphasizes flexibility through repeated development cycles that prioritize continuous software improvement and evaluation. Its core philosophy, as outlined in the Agile Manifesto, highlights values such as prioritizing individuals and interactions over processes and tools, working software over comprehensive documentation, customer collaboration over contract negotiation, and responsiveness to change over adherence to predetermined plans \cite{dora2013software} \cite{mehta2023agile}.
\begin{figure} [htb!]
    \centering
    \includegraphics[width=0.5\textwidth]{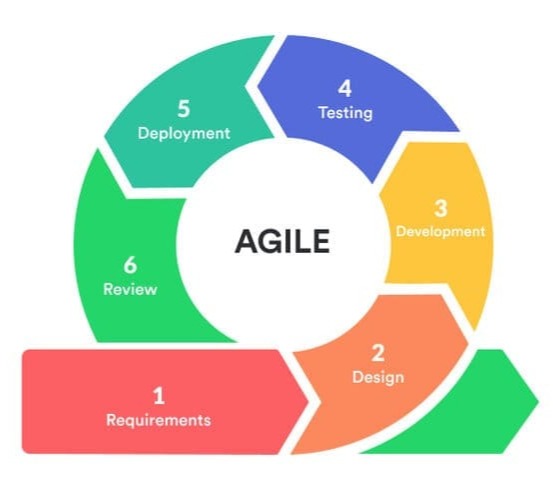}
    \caption{Agile Methodology}
    \label{fig:enter-label}
\end{figure}

Agile offers a set of advantages and limitations and has become one of the most widely adopted methodologies in modern information system development \cite{nova2022analisis}. Its primary strength lies in its adaptability to evolving requirements that may arise throughout the development process \cite{hidayah2024penerapan}. This flexibility enables development teams to continuously refine the product in accordance with the changing needs of users. Additionally, Agile emphasizes iterative cycles consisting of repeated stages requirements, design, development, testing, deployment, and review allowing each cycle to deliver measurable improvements. Another key advantage is the continuous feedback loop from stakeholders, which helps reduce the risk of deviations from user needs. Nonetheless, Agile is not without drawbacks; documentation produced under this methodology tends to be less detailed compared to traditional models, and its success requires strong commitment and involvement from all team members. Without proper management, projects can easily lose direction, resulting in scope creep and inefficiency.

The Agile methodology is highly suitable for the development of a Qibla compass application due to the nature of the system, which requires ongoing refinement and adjustments. Such an application integrates various complex components, including device sensors, spherical trigonometric algorithms, and an intuitive user interface. Through Agile, requirement gathering can be performed incrementally as new needs emerge, while the design phase can be adapted whenever revisions to the interface or application flow are required. Agile also facilitates a development process that prioritizes the creation of core features such as the compass system and the implementation of spherical trigonometry algorithms before extending to secondary functionalities such as the main menu and additional user tools. This flexibility demonstrates the strong compatibility of the Agile approach with the development of the system, as it allows developers to update or add features without disrupting the existing workflow.

The implementation of the Agile SDLC in the development of the Qibla compass application follows a structured series of iterative steps. As illustrated in Figure 1, these stages form a continuous cycle that supports incremental refinement throughout development. The methodology comprises several phases, including Requirements, Design, Development, Testing, Deployment, and Review \cite{pertiwi2023perancangan}. The first phase, Requirements, involves analyzing the functional and non-functional needs of the application, forming the foundational direction for the development process. The second phase, Design, is carried out using Figma to construct the interface layout and application interaction flow. The third phase, Development, focuses on building the application using Flutter and integrating the spherical trigonometry algorithm. This is followed by the Testing phase, which evaluates the functional behavior of the application including functional testing to verify that all features operate correctly. The subsequent phase, Deployment, involves releasing the application so it can be installed and utilized on user devices. The final phase, Review, serves as an evaluation stage to assess the overall development process and to determine necessary improvements or future iterations.

\subsection{Flutter}

Flutter is an open-source framework developed by Google for building mobile applications with a native-like experience from a single codebase. Unlike other hybrid frameworks, Flutter uses Dart as its programming language and adopts a unique approach in rendering its user interface. Flutter works by compiling Dart code into native ARM code for both iOS and Android, resulting in performance comparable to fully native applications \cite{ameen2022developing} \cite{mozharovskii2024performance}. Technically, Flutter operates based on the concept of “widgets,” where every interface component—from simple text elements to complex layouts is constructed as a widget that can be arranged hierarchically. The Flutter engine uses Skia as its graphics engine to render these widgets directly onto the canvas, without relying on a bridge to native components like many other hybrid frameworks.

The main advantage of Flutter lies in its ability to “write once, run anywhere” with truly native results. The hot reload feature allows developers to instantly view code changes without losing the application state, which greatly speeds up the development process  \cite{windmill2020flutter}. Flutter also offers consistent UI across platforms and full access to native features through platform channels  \cite{tran2020flutter}.However, Flutter also has several drawbacks, including larger application sizes compared to native apps, a third-party library ecosystem that is still smaller than native environments, and a learning curve for Dart, which may be unfamiliar to some developers.

The use of Flutter in developing a compass application, particularly a Qibla compass, provides several strategic advantages. First, Flutter’s ability to access platform-specific APIs allows the application to optimally utilize device sensors such as the magnetometer, GPS, and accelerometer for accurate Qibla direction calculations. Second, the hot reload feature is highly useful for developing and testing spherical trigonometry algorithms in real-time, enabling rapid iteration in refining calculation accuracy. Third, Flutter’s rich animation capabilities make it possible to create smooth and responsive compass visualizations \cite{tadas2024campus}.

In its implementation, Flutter is used to build the entire architecture of the Qibla compass application, from the presentation layer to the business logic. The presentation layer is built using Material Design and Cupertino widgets to ensure consistent appearance on both platforms. Sensor plugins such as \textit{sensors} and \textit{geolocator} are used to access magnetometer and GPS data, while custom painters are developed to render a responsive digital compass interface. The spherical trigonometry algorithm is implemented as separate Dart classes that can be tested independently. Platform channels are utilized to optimize sensor access performance at the native level, ensuring real-time data readings for precise Qibla direction calculations. Through this approach, the application is able to deliver an optimal user experience while maintaining the calculation accuracy that forms the core functionality.

\subsection{Testing}

\begin{enumerate}
    \item \textbf{Functional Testing}
    
    Functional testing is a type of software testing that focuses on verifying the functions of an application according to the requirements that have been established. This testing is carried out to ensure that each feature of the application behaves as expected by examining various input scenarios and validating the resulting output. In functional testing, testers do not consider the internal code or the structure of the application; instead, they only check whether the application provides the correct responses to user actions.

    The importance of functional testing lies in its ability to ensure that the application meets its functional specifications and is reliable for everyday use. Without adequate functional testing, an application is at risk of containing critical bugs that may disrupt the user experience or even cause system failure. This type of testing helps identify discrepancies between the actual behavior of the application and user expectations, thereby minimizing the risk of product rejection in the market. Moreover, functional testing plays a role in maintaining product quality consistently throughout the development cycle, ensuring that each new feature addition or code modification does not interfere with the functions that were already working correctly.

    Functional testing is particularly suitable for a Qibla compass application because the nature of the system relies heavily on very specific core functions. This testing will verify the correctness of the implementation of spherical trigonometry algorithms in calculating the Qibla direction, the accuracy of magnetometer and GPS sensor readings, and the precision of directional visualization on the compass interface. Test cases can be designed to examine various scenarios, such as changes in the Qibla direction when the user moves to a different location, the application’s response to sensor disturbances, and the accuracy of magnetometer calibration. Through functional testing, we can ensure that the Qibla compass application is truly reliable for Muslims in determining the prayer direction with sufficient precision, thereby fulfilling its primary purpose as an accurate and trustworthy worship support tool.

    \item \textbf{Performance Testing}
    
    Performance testing is a type of software testing that focuses on evaluating the performance of a system under certain workloads. This testing is conducted to measure non-functional attributes such as speed, responsiveness, stability, scalability, and application resource usage. Unlike functional testing, which verifies the correctness of features, performance testing examines how well the application operates under various conditions. Several forms of performance testing include load testing (testing under expected load), stress testing (testing beyond normal capacity), endurance testing (testing over extended duration), and spike testing (testing against sudden load surges).

    The importance of performance testing lies in its ability to prevent system failures in production environments, which can harm users and damage the product’s reputation. Without adequate performance testing, applications are at risk of experiencing slowdowns, crashes, or instability when used in real-world scenarios. Furthermore, performance testing ensures a smooth and responsive user experience, which is a critical factor in user retention and overall satisfaction.

    Performance testing is highly crucial for a Qibla compass application due to its real-time and continuous processing nature. This testing will measure the responsiveness of the application in updating the Qibla direction as the user moves, the efficiency of the spherical trigonometry algorithm in performing calculations, and the impact of continuous sensor usage on battery consumption.

    \item \textbf{Usability Testing}
    
    Usability testing is an evaluation method that focuses on the user experience when interacting with an application. Unlike functional testing, which validates feature correctness, usability testing measures aspects such as ease of use, efficiency, memorability, error rate, and users’ subjective satisfaction. This testing involves real users or representative users who are asked to complete specific tasks while observers record their behavior, difficulties, and feedback.

The importance of usability testing lies in its ability to bridge the gap between developers’ perceptions and the actual needs of users. Developers are often too familiar with the applications they create, making it difficult for them to identify the challenges that new users may encounter. Without usability testing, applications risk having a steep learning curve, confusing workflows, or unintuitive designs even if all features function correctly.

Usability testing becomes highly critical for a Qibla compass application because it will be used by diverse groups of Muslims across varying ages and technological backgrounds. This testing will examine the ease of sensor calibration for beginner users, the clarity of Qibla direction visualization under different lighting conditions, and the intuitiveness of the interface for elderly users who may be less familiar with modern technology.
    \end{enumerate}

\subsection{Spherical Trigonometry Algorithm}

Spherical trigonometry algorithms are mathematical methods used to calculate the relationship between points on the surface of a sphere based on the concept of spherical triangles. In this algorithm, the sides of the triangle are represented by great circles and the angles are measured at the center of the sphere. This approach forms the basis for various geographical calculations involving the Earth's surface, including determining the direction of the qibla, because the Earth is considered to be a nearly perfect sphere in geodesy \cite{AlAzzam2021}.

In this study, the spherical trigonometry algorithm was used to calculate the direction angle between the user's position and the position of the Kaaba in Mecca. The principle is to determine the shortest arc on the earth's surface that connects the two points. This arc is called a great circle path, which is the shortest path between two points on the surface of a sphere. By utilizing spherical trigonometry formulas, the direction of the qibla can be calculated accurately based on the user's latitude and longitude data.

\begin{enumerate}
\item \textbf{Concepts and Principles of Calculation}\\
In calculating the direction of the qibla, the earth is represented as a sphere with two important points, namely the user's location and the location of the Kaaba. These two points form a spherical triangle with the North Pole as the reference point. The angle formed at the user's location is the direction of the qibla that is being sought.

The spherical trigonometry formula used to determine the direction of the qibla (azimuth) is as follows:

\begin{equation}
\tan(\theta) = \frac{\sin(\Delta\lambda)}
{\cos(\phi_1)\tan(\phi_2) - \sin(\phi_1)\cos(\Delta\lambda)}
\end{equation}

where:\\
$\phi_1$ = user location latitude,\\
$\lambda_1$ = user location latitude,\\
$\phi_2$ = Kaaba latitude ($21.4225^\circ$ LU),\\
$\lambda_2$ = Kaaba longitude ($39.8262^\circ$ BT),\\
$\Delta\lambda = \lambda_2 - \lambda_1$.\\

This formula is derived from the basic laws of spherical trigonometry, which allows the relationship between three points on the Earth's surface to be calculated accurately. Since the position of the Kaaba is fixed, the main variable that affects the calculation results is the user's location coordinate \cite{Rahman2022}.

\item \textbf{Stages of Algorithm Calculation}\\
 Spherical trigonometric algorithms are mathematical methods used to calculate the relationships between points on the surface of a sphere based on the concept of spherical triangles. In this algorithm concept, the sides of the triangle are represented by great circles and the angles are measured at the center of the sphere. This principle forms the basis for calculating the direction of the qibla because the position of the Kaaba and the user's location are both located on the surface of the earth, which is close to a perfect sphere.

The steps for calculating the spherical trigonometry algorithm in the Qibla direction finder application begin with determining two main coordinates, namely the user's location coordinates (latitude and longitude) and the coordinates of the Kaaba in Mecca. Once both coordinates are known, the process continues by applying spherical trigonometry formulas to calculate the angular distance and azimuth (qibla direction) between the two points on the earth's surface.

The main steps in this algorithm include calculating the longitude difference between the user's location and the Kaaba, then using the sine, cosine, and tangent functions to obtain the initial azimuth value. This value is then corrected to be within the range of 0°–360°, which indicates the actual direction relative to true north. The direction of the qibla is then expressed in degrees, which can be used by the application to display the orientation of a digital compass.

The spherical trigonometry approach is considered more accurate than the planar method because it mathematically accounts for the curvature of the Earth. This method also has advantages in terms of the stability of calculation results at various latitudes and longitudes, so that it continues to provide accurate results in both tropical and high latitude areas. In addition, the accuracy of qibla direction measurements with this algorithm can also be improved with the use of modern geospatial data such as the World Geodetic System (WGS84), which is often integrated into GPS-based navigation systems \cite{Iqbal2022}.

\item \textbf{Accuracy of Calculations}\\
Calculation accuracy is a major factor in the application of spherical trigonometry algorithms to determine the direction of the qibla accurately. This algorithm is based on the mathematical relationship between two points on the Earth's surface connected by a great circle arc. Since the Earth is approximately spherical, the use of spherical trigonometry provides more realistic results than planar methods that assume the Earth is flat \cite{Hakim2021}.

The accuracy of qibla direction calculations is greatly influenced by the precision of input data in the form of latitude and longitude. Even a small error of 0.01° in the coordinate data can result in a difference of several degrees in the final result. In addition to coordinate factors, accuracy is also influenced by the mathematical formula used. Accuracy test results from various studies show that the spherical trigonometry method is capable of providing an error rate of less than 1°, even in extreme geographical conditions such as high latitudes. This makes the Spherical Trigonometry Algorithm very efficient for digital qibla direction applications, both on mobile devices and web-based systems. 

Considering these factors, it can be concluded that the accuracy of calculations using spherical trigonometric algorithms is highly dependent on the quality of input data, the geodetic model used, and the orientation correction mechanism on the device. The optimal combination of these three aspects can produce a precise, stable, and reliable qibla direction indicator system in various geographical conditions \cite{Maulana2022}.

\end{enumerate}

\section{Result and Discussion} \label{sec:result}

\subsection{System Implementation}

Based on the development stages that have been carried out, we successfully implemented the spherical trigonometry algorithm into a Flutter-based mobile application named QiblatKita. This application is designed as a digital solution to assist Muslims in accurately determining the Qibla direction in real-time. System implementation encompasses the full integration between software components (the calculation algorithm) and hardware (smartphone sensors) working synergistically. The QiblatKita application has undergone a comprehensive series of development processes, ranging from architectural design, coding, testing, to performance optimization, resulting in a stable and ready-to-use final product.

In the initial implementation phase, the development of the QiblatKita application faced various significant technical challenges. We encountered numerous bugs and system errors that disrupted the smooth operation of the application, with the main issues centering on sensor integration and calculation accuracy. Various technical obstacles arose, including inconsistencies in magnetometer data readings, the appearance of unnecessary error displays, and instability in the digital compass view.

Through a systematic debugging approach and repeated iterative testing, every identified issue was successfully resolved step-by-step. This continuous refinement process ultimately ensured that all system components could integrate and cooperate harmoniously, overcoming the technical constraints present in the early development stages.

The implementation of the spherical trigonometry algorithm is the core of the QiblatKita application, where we integrated the Qibla direction calculation formulas with real-time smartphone sensor readings. Initially, the system experienced precision issues in calculations due to sensor data noise and the limitations of mobile device processing power. However, through code optimization and the implementation of digital filters, we managed to significantly enhance calculation accuracy. The sensor calibration process, which was initially problematic, has also been perfected by adding a guided calibration feature that helps users correctly calibrate their devices before using the application.

\begin{figure} [htb!]
    \centering
    \includegraphics[width=0.60\linewidth]{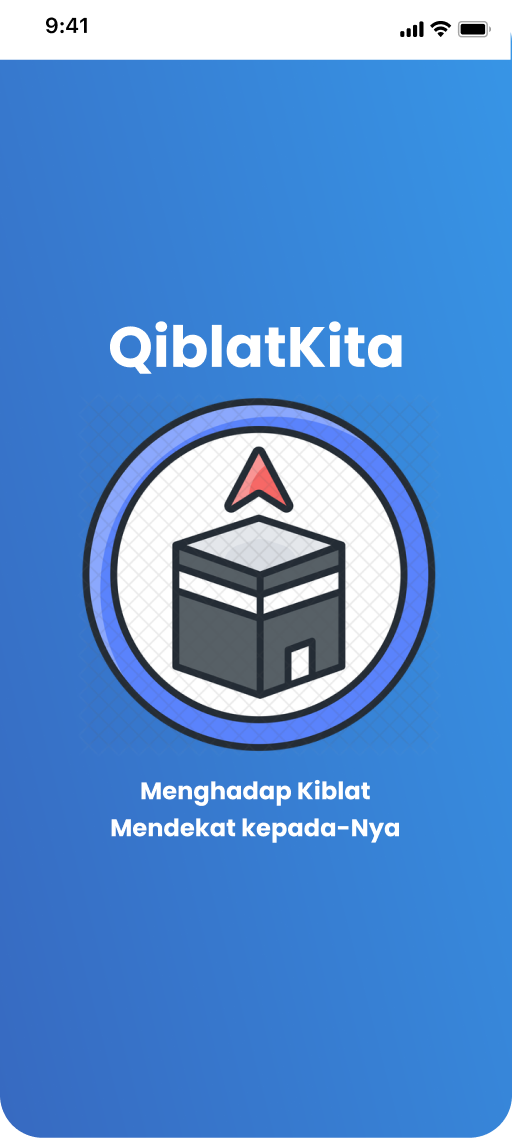}
    \caption{Opening Page Display}
    \label{fig:enter-label}
\end{figure}

\subsubsection{User Interface and Workflow} 

First and foremost, when users open the QiblatKita application, they are greeted by a splash screen designed with principles of user-centricity and simplicity. The visual display on this initial screen immediately communicates the identity and main purpose of the application, which is to serve as a dedicated digital aid for Muslims in accurately determining the Qibla direction for prayer (salat). The interface design is deliberately kept from being overly complex or cluttered with confusing elements; it only displays the application's logo, the name QiblatKita, and a tagline or icons that universally represent the function of a compass and worship. This provides a clear, professional, and easily understood impression for users of all backgrounds from the very first moment they launch the application.

Apart from welcoming the user, the presence of the splash screen in the QiblatKita application holds crucial technical and psychological roles. Technically, the brief moment the splash screen is displayed is used by the system to perform vital initializations, such as loading the GPS location detection library, preparing the magnetometer and accelerometer sensors, and compiling the application's main components that require a longer time to load. Psychologically, a splash screen with optimal display time creates an impression of responsiveness and professionalism, preventing the perception that the application has hung or crashed while the loading process is occurring in the background. For users, this brief pause also serves as a smooth transition from the operating system environment into the application's ecosystem, while simultaneously reinforcing the branding and visual identity of QiblatKita in the user's mind, thereby building early trust in the application's reliability as a dependable Qibla direction indicator.

\begin{figure} [htb!]
    \centering
    \includegraphics[width=0.60\linewidth]{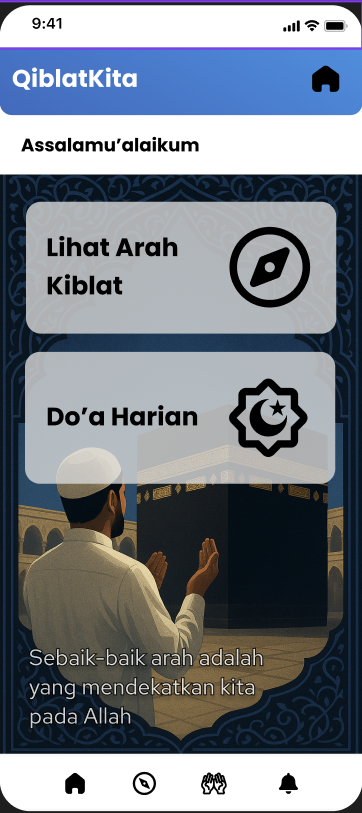}
    \caption{Main Menu Page Display}
    \label{fig:enter-label}
\end{figure}

The main menu page of the QiblatKita application is designed with a strong and authentic Islamic nuance to reinforce its identity as a spiritual-supporting application. The visual elements were carefully selected, including elegant Islamic calligraphy, a combination of green and gold colors often associated with Islamic culture, and characteristic geometric patterns from Islamic architecture adorning the background. These ornaments not only create a pleasing aesthetic but also immediately communicate the application's values and purpose from the moment the user opens it. Every detail is designed to create a calm and solemn atmosphere, reminding users that this is a spiritual tool, while building confidence that the application was developed with an understanding of and respect for Islamic values.

On the main menu page, users are presented with two primary function options displayed in the form of clear, easily accessible menu cards or large buttons. The first and most prominent option is``Qibla Compass'', which is the core feature of the application, directing users immediately to the digital compass interface for determining the prayer direction. The second option is ``Collection of Prayers (Do'a)'' which serves as a practical guide containing daily prayers and specific prayers related to salat (prayer), making it easier for users to perform ``dhikr'' and supplication after performing their worship. For smoother navigation, a simple and intuitive Navigation Bar is placed at the bottom of the screen, allowing users to quickly switch between the main page, the compass page, the prayer page, and the settings page without having to return to the main menu first.

\begin{figure} [htb!]
    \centering
    \includegraphics[width=0.60\linewidth]{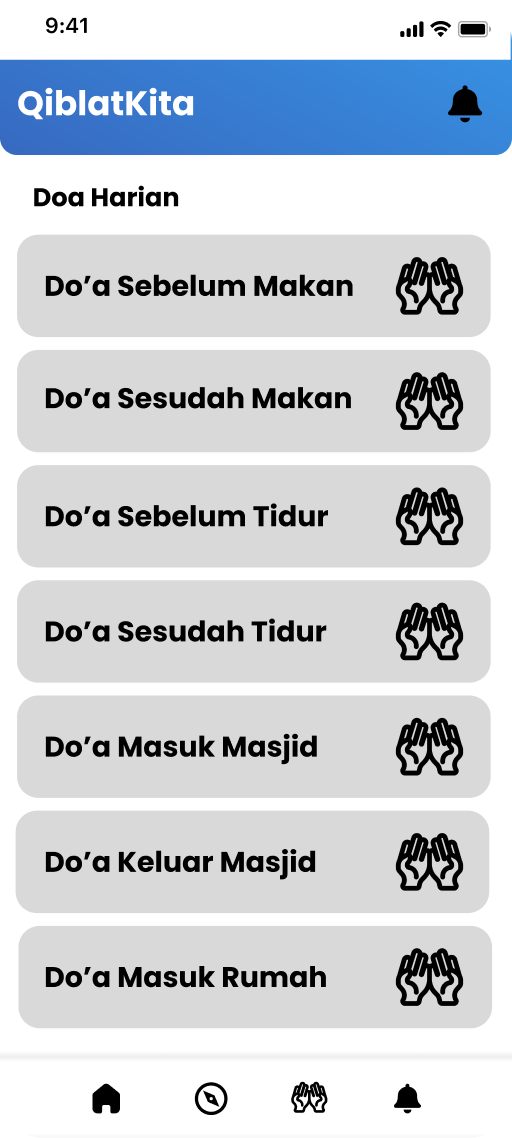}
    \caption{Daily Prayers Page Display}
    \label{fig:enter-label}
\end{figure}

The ``Collection of Prayers (Kumpulan Doa)'' page in the QiblatKita application is designed as a comprehensive yet easily accessible spiritual guide, displaying a neatly organized list view interface containing a collection of essential daily prayers for a Muslim. Each prayer is presented as a clear list item, showing the prayer title in Indonesian. The design is clean and spacious, using easy-to-read typography and complemented by simple icons next to each prayer title, providing an aesthetic touch and aiding the user in visual navigation. When a user taps on one of the prayers from the list, the application transitions to a detail page that displays the complete text of the prayer along with its translation, allowing users to comfortably study, read, and memorize these prayers without distraction.
\begin{figure}[htb!]
    \centering
    \includegraphics[width=0.60\linewidth]{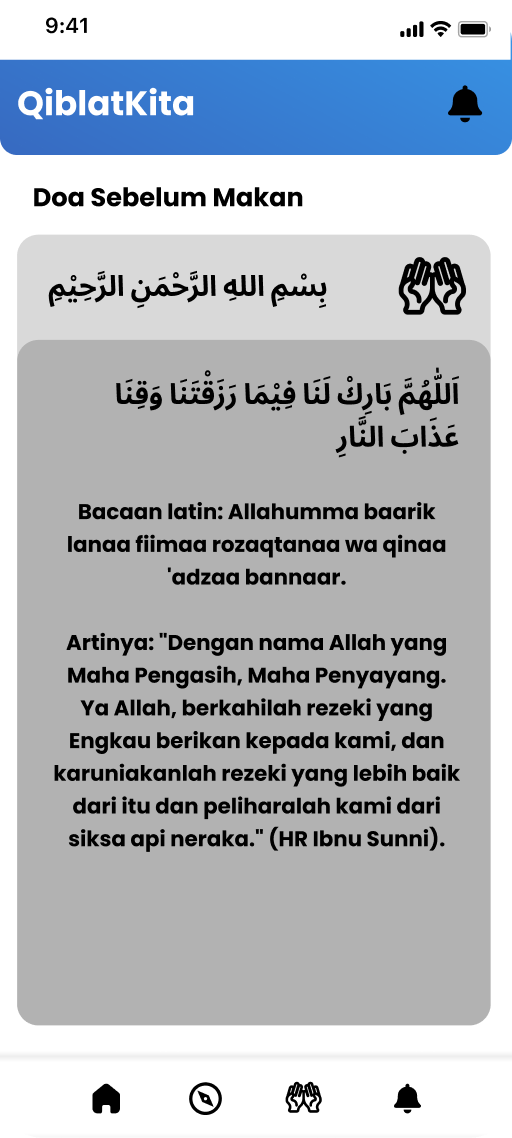}
    \caption{Daily Prayers Page Display}
    \label{fig:enter-label}
\end{figure}
When the user taps on one of the available prayers from the list, the application immediately transitions to the prayer detail page, which displays the complete content in a structured and easy-to-learn format. This page presents the entire prayer text in three forms: the original text in Arabic with a clear and legible font, accompanied by Latin transliteration beneath it as a pronunciation guide for users who are not yet fluent in reading Arabic script, and a complete Indonesian translation placed at the bottom to ensure understanding of the meaning and context of the prayer being recited.
\begin{figure}[htb!]
    \centering
    \includegraphics[width=0.60\linewidth]{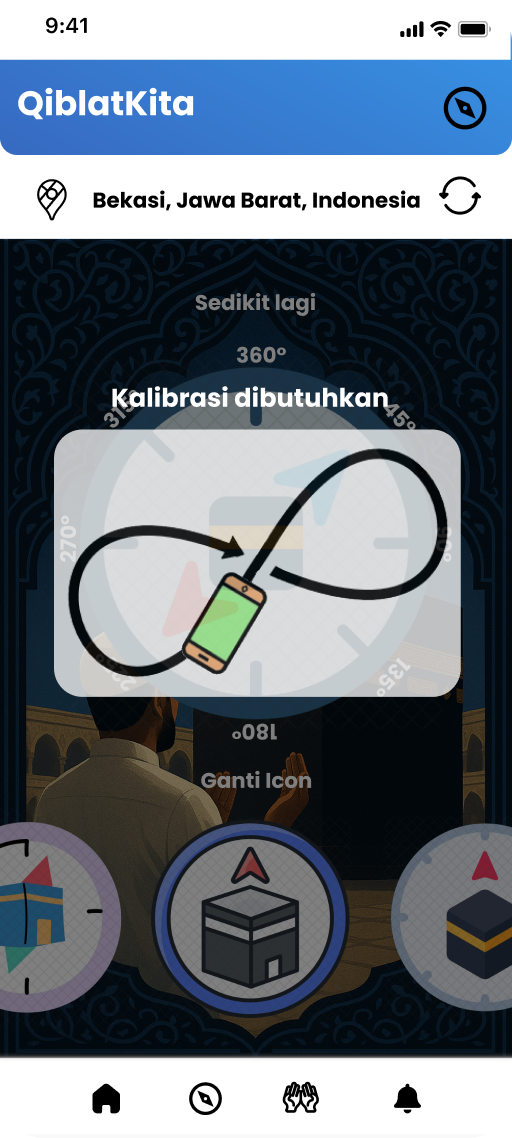}
    \caption{Compass Calibration Page Display}
    \label{fig:enter-label}
\end{figure}
When the user presses the $**"View Qibla Direction"**$ button on the main menu, the application immediately transitions to the Qibla compass page, which begins with a clear instructional overlay. This page automatically displays an important notification guiding the user to perform device calibration first. The instructions prompt the user to move their device in a 180-degree rotational pattern several times, a standard technique for magnetometer sensor calibration. This process is designed to align and calibrate the smartphone's internal magnetometer sensor to ensure reliable direction readings and minimize magnetic field interference from the surrounding environment, which is a critical step for supporting system accuracy.
\begin{figure}[htb!]
    \centering
    \includegraphics[width=0.60\linewidth]{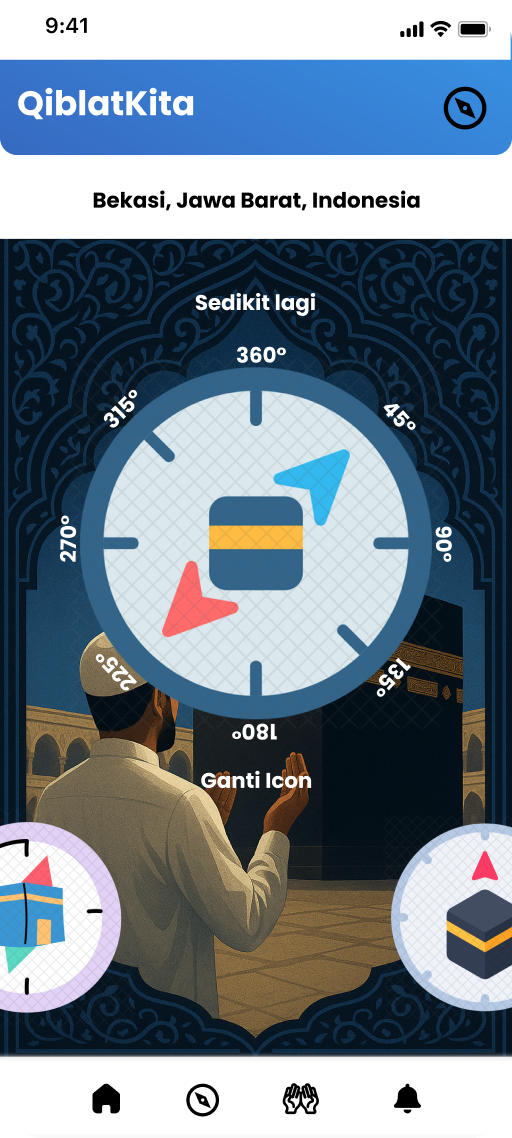}
    \caption{Qibla Compass Page Display}
    \label{fig:enter-label}
\end{figure}
Once the calibration process is complete, the Qibla compass page will display the digital compass interface, which immediately becomes active, working to find and indicate the Qibla position. This compass view is designed with clear and intuitive visuals, featuring a main pointer needle that moves in real-time, pointing toward the Ka'bah based on the user's geographical location. A key advantage of this page is the ``compass view customization feature'', where users can select various themes and compass styles according to their personal preference, ranging from a classic display with a traditional design, a modern view with minimalist elements, to a high contrast display that is easy to see in various lighting conditions. This flexibility ensures visual comfort for users with diverse tastes.

In addition to displaying the general direction, the application is also equipped with a ``precision feature'' that provides detailed information on how closely the indicated direction aligns with the actual Qibla position. This feature is displayed in the form of text indicating how close our direction is to the Qibla, accompanied by numerical information on the degree of deviation from the ideal Qibla direction. Some display variants even include a ''precision mode`` feature that provides step-by-step guidance to perfect the direction, such as instructions like ``turn slightly left'' or ''shift slowly right`` until optimal accuracy is achieved. This combination of a responsive digital compass, display customization options, and precision features makes the experience of determining the Qibla direction not only accurate but also interactive and informative for the user.

\subsection{Testing}

Based on a comprehensive evaluation, the QiblatKita application has undergone a series of rigorous tests to ensure product quality and reliability. ``Functional testing'' was conducted to verify that every component of the application operates according to the established specifications. The results show that all features and navigation buttons function perfectly as expected, ranging from sensor calibration, the digital compass display, to the theme customization feature. Most importantly, the implementation of the spherical trigonometry algorithm as the core of the Qibla direction calculation proved to have a sufficiently high level of accuracy in directing the compass towards the Ka'bah coordinates, with a negligible deviation for the practical purpose of determining the prayer direction.

Furthermore, performance testing was carried out to analyze the stability and responsiveness of the application under various conditions. The test results proved that the application displays stable and optimal performance, indicated by the absence of crashes, force close incidents, or hangs during the intensive testing period. The application was also able to maintain consistency in sensor readings and real-time display updates without causing significant battery drain. The processing of the spherical trigonometry algorithm data ran efficiently in the background without burdening the device's CPU, ensuring smooth operation even on mid-range specification devices.

The aspect of usability or ease of use was also the focus of in depth testing. The QiblatKita application proved to be highly user-friendly for people from various backgrounds, thanks to its simple, intuitive, and free of unnecessary complexity interface. The structured navigation design allows users even those who are less tech-savvy to easily access the main features in just a few taps. The results of usability testing on a sample of users with varying ages and technological backgrounds showed a high level of satisfaction, with the majority of respondents stating they could determine the

\subsection{Discussion}

Based on the entire research workflow that has been conducted, it can be discussed that the implementation of the spherical trigonometry algorithm in the QiblatKita application is proven to be effective and reliable. The high level of accuracy in determining the Qibla direction demonstrates that the mathematical approach based on spherical trigonometry is a suitable computational solution for this kind of spiritual geolocation problem. Performance test results showing application stability without crashes or forced closes indicate that the integration between the complex algorithm, hardware sensor readings, and the Flutter framework has been well optimized. This success is inseparable from the Agile SDLC development approach, which allowed for continuous iteration and improvement, especially in overcoming early technical challenges such as sensor reading inconsistencies and power consumption optimization. These findings strengthen the proposition that scientifically-based digital solutions can be reliable in supporting daily worship activities.

From the user's perspective, the success of the QiblatKita application is not only measured by technical accuracy but also by its ease of use and the meaningfulness of its function in religious life. The intuitive interface design and simple navigation successfully accommodate the needs of users from various age backgrounds and levels of technological literacy. The presence of supporting features such as the collection of prayers (do'a) not only adds value to the application but also creates a comprehensive digital ecosystem for Muslim spirituality. These results are in line with findings from previous studies that emphasize the importance of the usability aspect in religious applications. Practically, this research has contributed to demonstrating how the integration between classical astronomy ($*ilm falak*$), modern technology, and user-centered design principles can produce a solution that is not only technically accurate but also easily adopted and directly beneficial to the community in fulfilling their religious daily practice needs.

\section{Conclusion} \label{sec:conclusion}

Based on the entire research and development process that has been carried out, it can be concluded that the QiblatKita application, built with the implementation of spherical trigonometry algorithms on the Flutter platform, has successfully become an accurate, stable, and easily accessible digital solution for determining the direction of the qibla. Functional, performance, and usability testing results prove that the application not only meets technical criteria with high calculation accuracy and reliable performance without crashing, but also successfully presents an intuitive interface that can be used by various Muslim communities. Thus, this research has contributed to bridging classical astronomy with modern technology, providing a practical and reliable tool to support the accuracy of prayer in accordance with Islamic law.

\section*{Acknowledgment}
The author's wishes to acknowledge the Informatics Department UIN Sunan Gunung Djati Bandung, which partially supports this research work.

\bibliographystyle{./IEEEtran}
\bibliography{./IEEEabrv,./IEEEkelompok1}


\end{document}